\documentclass[10pt]{article}

\usepackage{graphicx}
\usepackage{dcolumn}
\usepackage{bm}
\usepackage{amsmath}
\usepackage{amssymb}
\begin{document}
\begin{center}
{{\LARGE \bf Heavy dilepton in nucleus nucleus collision at LHC
energy }} \\
\bigskip
{\large \bf Sarbani Majumder$^*$\footnote{email: sarbanimajumder@gmail.com}} \\
{$^*$Department of Physics, Bose Institute, Kolkata, India} 
\bigskip
\end{center}

\begin{abstract}
We present a study of $\tau^+ \tau^-$ lepton pair production in 
Pb + Pb collisions at $\sqrt{s_{NN}}$ = 5.5 TeV. The larger
mass of tau lepton compared to electron and muon leads to 
considerably small hadronic contribution to the
$\tau^+ \tau^-$ pair invariant mass (M) distribution relative
to the production from thermal partonic sources. The
quark–anti-quark annihilation processes via intermediary
virtual photon, Z and Higgs bosons have been
considered for the tau lepton production.
The contribution from Drell–Yan process is found to dominate
over thermal yield for $\tau^+ \tau^-$ pair mass from 4 to
20 GeV at the LHC energy. 
We also present the ratio of
$\tau$ lepton pair yields for nucleus–nucleus collisions
relative to yields from p + p collisions scaled by number
of binary collisions at LHC energies as a function $\tau$
lepton pair inavariant mass.
The ratio is found to be significantly above
unity for the mass range 4 to 6 GeV.
This indicates the possibility of detecting $\tau^+ \tau^-$
pair from quark–gluon plasma (QGP) in the mass window 4 to 6
GeV.
\end{abstract}


\section{Introduction}
The heavy dilepton pairs
namely $\tau^{+}\tau^{-}$ created in Pb-Pb collision at LHC
energy is discussed in this work. 
The major advantage of 
looking at $\tau^{+}\tau^{-}$ dilepton pair arises due 
to the mass of the  $\tau$ ($\sim$ 1.77 GeV).
The  $\tau$ pair mass distribution would then start 
beyond known contribution of hadronic
resonances ($\omega$, $\rho$ and $\phi$) which dominates
in the respective mass regions 
in $e^{+}e^{-}$ and $\mu^{+}\mu^{-}$  sector. 
This would in turn mean the remaining contribution
for $\tau$ production are due to thermal sources 
from partonic medium, pion annihilation in
hadronic medium and Drell Yan Mechanism. 
We think that the results will definitely provide some
useful baseline for experimental search and further 
detailed studies.
\section{Source of $\tau$ Lepton Pair in Heavy Ion Collision}
\begin{figure}[!h]
\begin{center}
\includegraphics[scale=0.2]{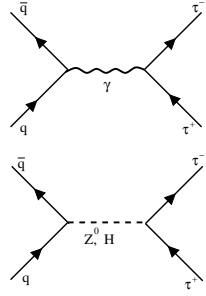}
\caption{Feynman diagrams for heavy dilepton production.}
\label{fig1}
\end{center}
\end{figure}

The main source of heavy dilepton $\tau^{\pm}$ pair production 
we have considered in this work is 
quark and anti-quark annihilation through photon, Z and Higgs bosons 
intermediated process. The corresponding Feynman diagrams are shown 
in Fig.~\ref{fig1}. They all contribute to 
thermal production of $\tau^{\pm}$ pair in quark gluon plasma.
The productions for these processes are evaluated from the matrix 
elements indicated below. The matrix element for the process 
$q\bar{q} \rightarrow {Z} \rightarrow \tau^+\tau^-$  is given by,
\begin{equation}
 M_Z =\frac{g^2}{4cos^2\theta_w}\frac{1}{(q^2-{m_z}^2)}[\bar{v(p_2)}
{\Gamma_q}u(p_1)][\bar{u(k_1)}{\Gamma_\tau}v(k_2)]
\end{equation}
where, 
\begin{equation*}
 \Gamma_q=\gamma^\mu({c_V}^q-{c_A}^q\gamma_5)
\end{equation*}
\begin{equation*}
\Gamma_\tau=[{\gamma_\mu}-\frac{{q_\mu}{\gamma_\nu}{q^\nu}}
{{m_z}^2}][{c_V}^\tau-{c_A}^\tau{\gamma_5}]
\end{equation*}
Finally putting the values of the parameters we have,
\begin{eqnarray}
 \lvert M_Z^2 \rvert&=&\frac{g^4}{16cos^4 \theta_w}
\Big[0.5742[(s+t-m_q^2-m_\tau^2)^2+
(m_q^2+m_\tau^2-t)^2\nonumber\\
&&+(s+t-m_q^2-m_\tau^2)^2
(1-\frac{s}{2m_Z^2})]
+s(0.858m_q^2+1.14m_\tau^2)\nonumber\\
&&-0.59m_q^2m_\tau^2
-0.0041\frac{s m_q^2 m_\tau^2}{m_Z^2}+0.002
\frac{s^2 m_q^2 m_\tau^2}{m_Z^2}\Big]
\label{eq.z}
\end{eqnarray}
Similarly, for the photon mediated process we have: 
\begin{equation}
 \lvert M_\gamma^2 \rvert=\frac{e^2 e_q^2}{s^2}
\Big[\frac{1}{4}[(m_q^2+m_\tau^2-t)^2
+(s+t-m_q^2-m_\tau^2)^2]+(\frac{s}{2}-m_q^2)m_\tau^2\Big]
\label{eq.photon}
\end{equation}
$e_q$ is the average charge of quarks, e is the electronic charge.
The matrix element of the interference term is 
$M_Z M_\gamma^*+M_Z^*M_\gamma$.
\begin{eqnarray}
&& M_Z M_\gamma^*+M_Z^*M_\gamma=\frac{g^2 e_q e}{4 cos^2 \theta_w}
\frac{1}{s(s-m_z^2)}\Big[(m_q^2+m_\tau^2-t)^2 \nonumber \\
&-&(s+t-m_q^2-m_\tau^2)^2[0.0912
+12(4-\frac{s}{m_Z^2})]
+0.0912\frac{s^3}{m_Z^2} \nonumber \\
&-&0.1824\frac{m_\tau^2}{m_Z^2}s^2
+0.5472\frac{m_q^2 m_\tau^2}{m_z^2}s
+0.1824 m_q^2 s-1.09 m_q^2 m_\tau^2\Big]
\label{eq.interference}
\end{eqnarray}
Finally, the matrix element for the Higgs mediated process is:
\begin{equation}
\lvert M_H^2 \rvert=\frac{m_q^2 m_\tau^2}{vev (s-m_H^2)^2}
[2m_q^2-\frac{s}{2}][2m_\tau^2-\frac{s}{2}]
\label{eq.H}
\end{equation}
\par
Here, $m_q$, $m_\tau$, $m_Z$, $m_H$, are the masses of quarks,
$\tau$ leptons, $Z$ boson and Higgs respectively.
($p_1$, $p_2$) and ($k_1$, $k_2$) 
are initial state and final state momenta respectively.
The total production cross section ($\sigma_q$) 
of $\tau^+\tau^-$ is obtained by taking a coherent sum
of the matrix elements given in Eqs.(\ref{eq.z}), (\ref{eq.photon}), 
(\ref{eq.interference}) and (\ref{eq.H}) with the following
values of various parameters:  
$m_Z$=91 GeV, $M_\tau$=1.78 GeV, $m_H$=120 GeV, $sin\theta_w$=0.234
$C_A^q$=0.5, $C_V^q$=0.19, $C_A^\tau$=-0.5, $C_V^\tau$=-0.03 and
Higgs vev=246 GeV.
\section{$\tau$ Lepton From Drell-Yan Process}
The total production cross-section $\sigma_q$ is folded
by the parton distribution functions to obtained
the $\tau$ lepton pair yield in p-p collisions.
In the present work CTEQ5M PDF \cite{cteq} have been taken 
to obtain this.
The DY production of $\tau$ lepton in Pb-Pb collision is; 
\begin{equation*}
\frac{dN}{dM^2dy}=\frac{N_{\mathrm coll}(b)}
{\sigma^{pp}_{\mathrm in}}\times
\frac{d\sigma^{pp}}{dM^2dy}
\end{equation*}
where, $N_{\mathrm coll}(b)$ is the number of binary
nucleon nucleon collisions an impact parameter b
calculated using Glauber model and $\sigma_{in}$ 
is the inelastic cross section for p-p interaction. We
have taken  $\sigma_{in}=60$ mb and $b=3.6$ fm 
corresponding to $0-5\%$ centrality at $\surd{s}_{NN} = 5.5$ 
TeV.
The shadowing of parton distribution functions 
has been taken from \cite{eskola}.
\section{Space-Time Evolution Of $\tau$ Lepton Pairs}
The space time evolution of the system formed
in Pb+Pb collisions at $\sqrt{s_{\mathrm {NN}}}$ = 5.5 TeV 
has been studied by using ideal relativistic hydrodynamics 
\cite{ideal} with longitudinal 
boost invariance \cite{bjorken} and cylindrical symmetry. 
Our assummption is that the system reaches equilibration at a time 
$\tau_{i}$ = 0.08 fm/c after the collision. 
The initial temperature, $T_{i}$ is considered to be 700 MeV and 
is calculated assuming the hadronic multiplicity (dN/dy) to
 be of the order of 2100 \cite{armesto}.
The equation of state (EOS)  obtained from the Lattice
QCD calculations by the MILC collaboration \cite{milc} for 
the partonic phase is used here. 
For the hadronic phase EOS all the resonances with mass 2.5 GeV
have been considered~\cite{hadronicph}.
The transition temperature ($T_{c}$) between hadronic phase and 
partonic phase is taken to be 175 MeV~\cite{tc}. 
We consider kinetic freeze out temperature, $T_{f}$ = 120 MeV.
In this context we want to mention that
the gluon fusion process was found to be dominant for mass
of lepton pair greater than the mass of W boson. Our results are
concentrated in the mass range of 4 to 20 GeV, where
the contribution from such process is found to be orders of
magnitude smaller compared to the rest of the sources of
$\tau^{\pm}$ pair production.
\section{Results}
\label{sc.result}

\begin{figure} [!htb]
\begin{center}
\includegraphics[scale=0.25]{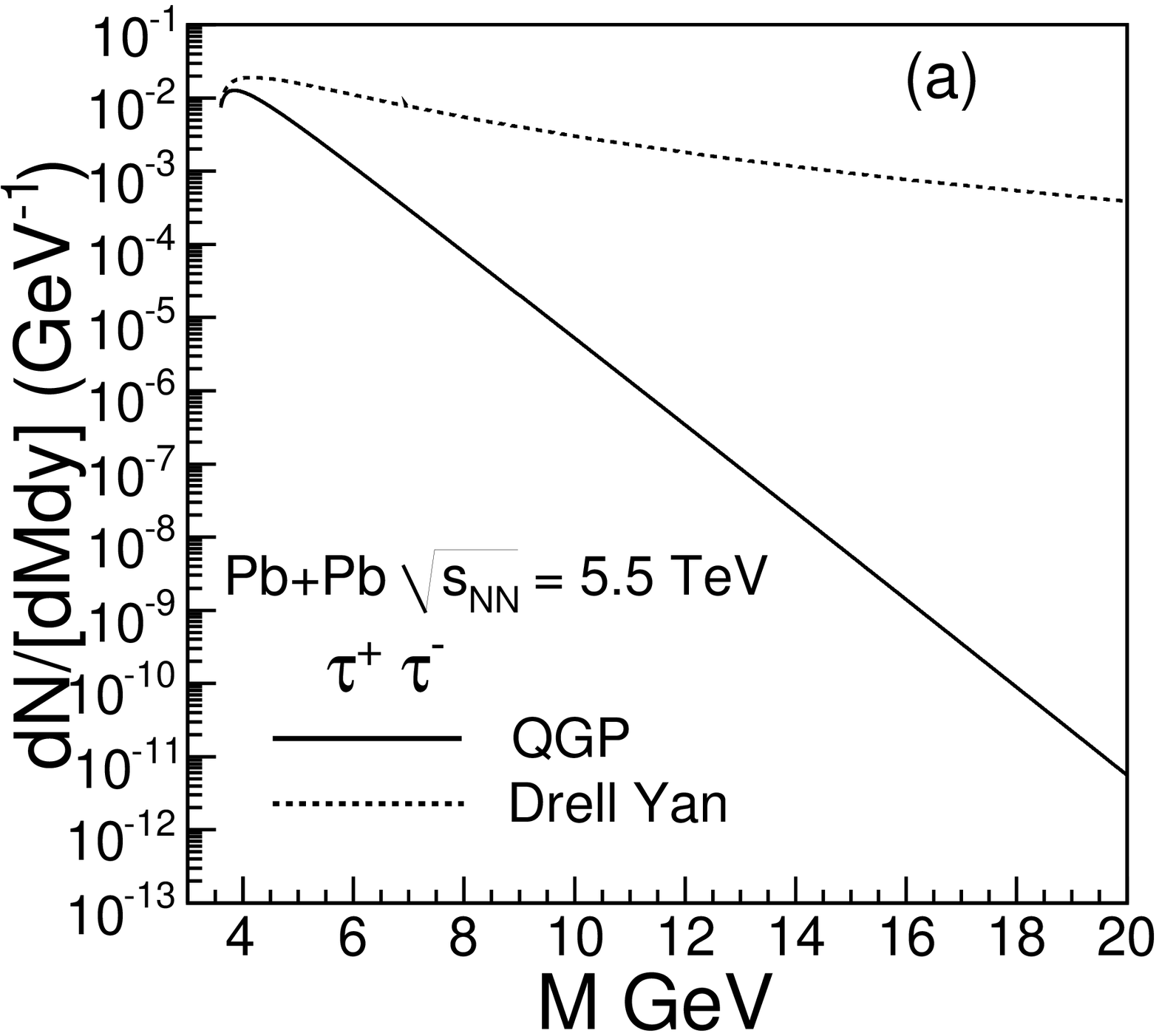}
\includegraphics[scale=0.285]{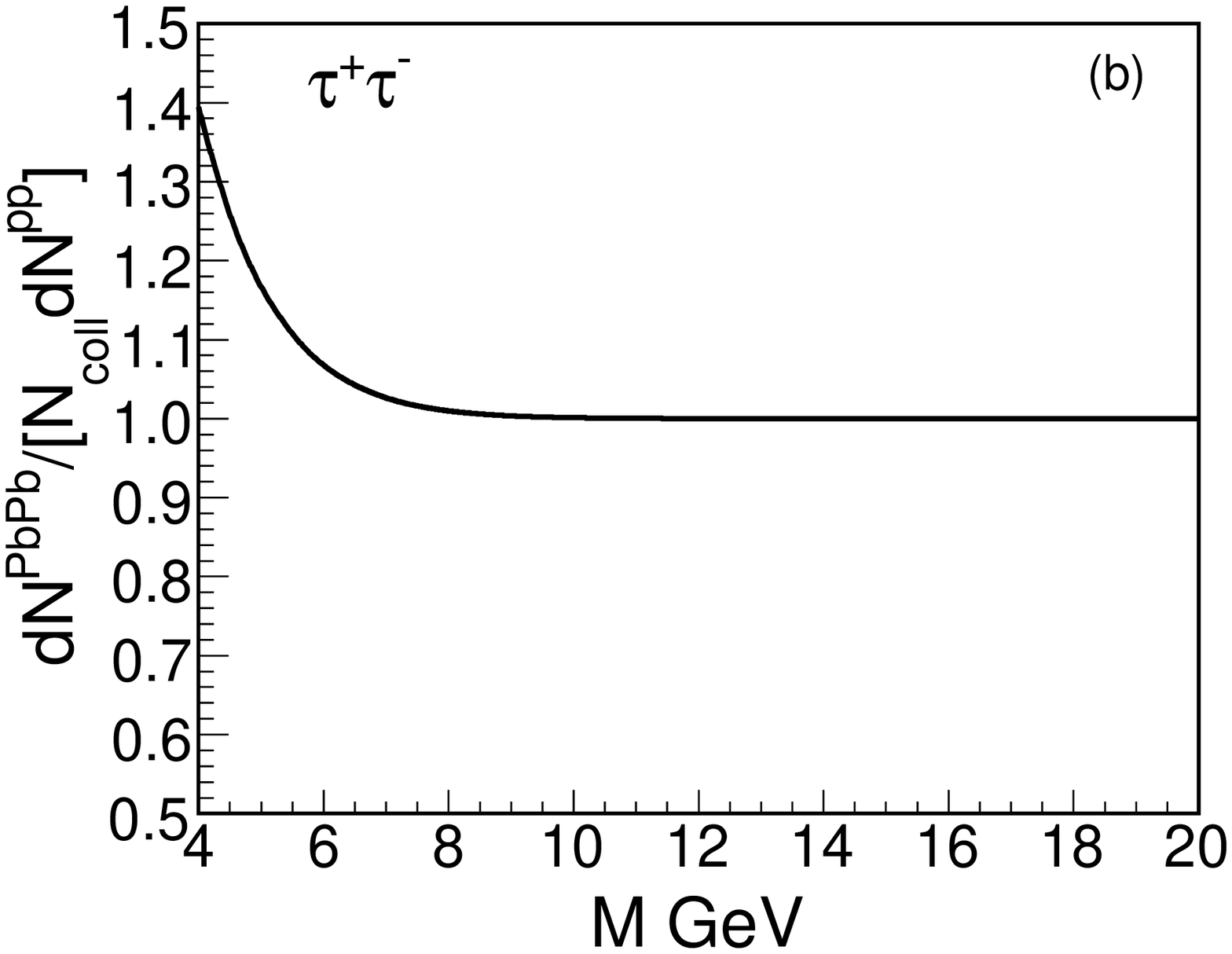}
\caption{(a) $\tau$ lepton pair yields as a function of 
invariant mass of the pair is displayed for  Pb+Pb collisions
at $\sqrt{s_{\mathrm {NN}}}$ = 5.5 TeV. 
Solid line indicates the spectra 
from quark gluon plasma and the dashed line stands for 
contribution from DY process.
In (b) the ratio 
$\frac{dN^{PbPb}}{dMdy}$ / $[N_{coll}\frac{dN^{pp}}{dMdy}]$
is shown,
here $\frac{dN^{PbPb}}{dMdy}$ is the sum of the contribution
shown in (a), 
$\frac{dN^{pp}}{dMdy}]$ is the DY  contribution from $pp$ collision,
and $N_{coll}$ =1369.477  for Pb+Pb collisions at 
$\sqrt{s_{\mathrm {NN}}}$ = 5.5 TeV.}
\label{fig2}
\end{center}
\end{figure}
 The yield ($\frac{dN}{dMdy}$) 
for $\tau$ dilepton pair as a function of $\tau^+\tau^-$ 
pair invariant mass for Pb+Pb collisions at 
$\sqrt{s_{\mathrm {NN}}}$ = 5.5 TeV is shown in 
figure \ref{fig2}(a).  
The contributions from
Drell Yan (DY, dashed line) and thermal partonic medium 
(QGP, solid line) are shown.
As evident from the figure, the Drell Yan contribution
is higher than the thermal contribution for all the mass 
range studied. The difference seems to increase with 
increase in $\tau^+\tau^-$ pair mass. 
\par
Figure \ref{fig2}(b) shows the ratio 
$\frac{dN^{PbPb}}{dMdy}$ / $[N_{coll}\frac{dN^{pp}}{dMdy}]$, where 
$\frac{dN^{PbPb}}{dMdy}$ is the sum of all the the contributions
 shown in  Figure~\ref{fig2}(a)
from Pb+Pb collisions. The $[N_{coll}\frac{dN^{pp}}{dMdy}]$, 
is the number of binary collisions
scaled contribution from DY process.
This contribution can be estimated from the  measurement in p+p
collisions at the same energy ($\sqrt{s}$ = 5.5 TeV).
If there is no QGP formation then the ratio should always be
equal to unity indicating the fact that the dilepton yield in
the nucleus-nucleus collision is the collection of individual
nucleon-nucleon collision only. However,
we observe that the ratio is above unity 
for the mass range of 4 to 6 GeV.
Starting with a value of 1.4 at mass of 4 GeV it decreases 
towards unity beyond mass of 6 GeV. 
This indicates that one should be able to extract a clear 
information of thermal contribution
from partonic source at LHC energies using heavy dilepton pair 
measurement within the mass window of 4 to 6 GeV. 
\section{Summary}
 A  study of 
$\tau$ dilepton pair production at LHC energy has been carried out
in this work. We have considered Pb+Pb collisions
at midrapidity for $\sqrt{s_{\mathrm {NN}}}$ = 5.5 TeV. 
It is expected that this energy 
should allow a significant production of $\tau$
leptons. The main motivation of considering the $\tau$
lepton is its mass.
Because of heavy mass the tau lepton yield is 
expected not to suffer from the huge background production. 
The main sources for  $\tau$ pair 
production considered here by quark and anti-quark annihilation 
mediated through photon, Z and Higgs boson. 
The contribution from pion annihilation 
process is few orders of magnitude small compared 
to both thermal and Drell Yan contributions. 
The Drell Yan contribution is found to be higher 
than the thermal contribution from partonic
sources for the entire mass range studied. 
The non-thermal contributions could be measured 
experimentally through p+p collisions, then the ratio of 
yields from nucleus-nucleus collisions 
to the yields for the binary collision scaled p+p collisions 
is found to be above unity for 
the mass range of 4-6 GeV. This indicates the window in mass 
region for $\tau$ dilepton pair 
where the thermal production can be studied at LHC energy using
heavy dilepton pairs as an observable. 
\section{Acknowledgement}
The support of Prof. Jan e Alam, VECC, Dr. B. Mohanty, NISER, 
Prof. Sanjay K. Ghosh and Dr. Rajarshi Ray of Bose Institute,
Kolkata is gratefully acknowledged.

\end{document}